\documentclass[prl,twocolumn,amsmath,amssymb]{revtex4-1} %

\usepackage{epsfig,amsmath}
\usepackage{subfigure}
\usepackage{graphicx}% Include figure files
\usepackage{dcolumn}% Align table columns on decimal point
\usepackage{stmaryrd}
\usepackage{mathrsfs}
\usepackage{pifont}
\usepackage{amsthm}
\usepackage{amssymb}
\usepackage{bm}
\usepackage{latexsym}
\usepackage{float}
\usepackage[colorlinks=true,linkcolor=blue,citecolor=blue]{hyperref}
\usepackage{color}

\theoremstyle{plain}

\def\pra#1{{ Phys.\ Rev. A\/} {\bf#1}}
\def\prb#1{{ Phys.\ Rev. B\/} {\bf#1}}

\def\prl#1{{ Phys.\ Rev.\ Lett.} {\bf#1}}

\def\sci#1{{ Science} {\bf#1}}

\def\rmp#1{{ Rev. \ Mod. \ Phys.} {\bf#1}}
\def\nat#1{{ Nature} {\bf#1}}

\def\njp#1{{ New. J. \ Phys.} {\bf#1}}

\begin{document}

\title{Simple Circuits for Exact Elimination of Leakage in a Qubit Embedded in a Three-level System}

\author{Yifan Sun$^{1,2,3}$, Junyi Zhang$^{1}$, Lian-Ao Wu$^{2,3}$}\thanks{Author to whom any correspondence should be addressed. Email address: lianao.wu@ehu.es}

\affiliation{$^{1}$State Key Laboratory of Magnetic Resonance and Atomic and Molecular Physics, Wuhan Institute of Physics and Mathematics, Chinese Academy of Sciences, Wuhan 430071, China \\ $^{2}$Department of Theoretical Physics and History of Science, The Basque Country University (EHU/UPV), PO Box 644, 48080 Bilbao, Spain \\ $^{3}$Ikerbasque, Basque Foundation for Science, 48011 Bilbao, Spain}

\date{\today}

\begin{abstract}
Leakage errors damage a qubit by coupling it to other levels. Over the years, several theoretical approaches to dealing with such errors have been developed based on perturbation arguments. Here we propose a different strategy: we use a sequence of finite rotation gates to exactly eliminate leakage errors.  The strategy is illustrated by the recently proposed charge quadrupole qubit in a triple quantum dot, where there are two logical states to support the qubit and one leakage state. %which is useful for describing a triple quantum dot that can protect logical qubits from uniform electric field fluctuations by generalizing the concept of a decoherence-free subspace (DFS), we exactly solve the problem of eliminating leakage errors in the single qubit rotating operations. 
We find an ${\it su}(2)$ subalgebra in the three-level system, and by using the subalgebra we show that ideal Pauli $x$ and $z$ rotations, which are universal for single-qubit gates, can be generated by two or three propagators of experimentally-available Hamiltonians. The proposed strategy does not require additional pulses, is independent of error magnitude, and potentially reduces experimental overheads. In addition, the magnitude of detuning fluctuation can be estimated based on the exact solution. %Our results suggest that single qubit operations in the system described by similar Hamiltonian can perfectly get rid of leakage errors, which is applicable to fault-tolerant quantum computing with solid-state elements. 
\end{abstract}


\maketitle
The physical realization of quantum computer poses an unprecedented challenge to our capabilities of controlling the dynamics of quantum systems. While there have been many attempts to overcome this challenge, the perfect controllability of semiconducting quantum dots makes them promising candidates for universal quantum computation~\cite{A1,A2,A3,A4,A5,A6}. A universal quantum computer is the ultimate information processor in modern quantum technology, which uses quantum bits (qubits) and quantum circuits to perform computations. A qubit consists of an idealized pair of orthonormal quantum states. However, this idealization neglects other states which are typically present and can mix with those supporting the qubit. Such mixing is termed as {\em leakage}. Leakage may be the result of the application of gate operations, or induced by system-bath interactions~\cite{WuBrumer09,A7,A8,A9,A10,A11,A12}. %Two ways that a qubit operation can fail to have high fidelity are (i) the qubit could decay or dephase during the imperfect operation and (ii) quantum information could leak out of the qubit's logical subspace into other quantum states in the physical system due to the coupling between them~\cite{A8,A9,A10,A11,A12}. While several recent proposals in quantum dots have focused on suppressing dephasing from environmental noise~\cite{A13,A14}, relatively little effort has gone into suppressing leakage~\cite{A12}. 
Several strategies for combating the leakage errors have been developed for different systems, in particular the semiconducting qubit setup which is the main subject of this work, including analytic pulse shaping~\cite{A9,A15} and optimal quantum control~\cite{A16,A17}.  Ref.~\cite{WuLidar02} also presents a general leakage-elimination method for removing such errors by using simple decoupling and recoupling pulse sequences of the leakage elimination operator (LEO). Nonperturbative LEO was recently introduced  for nonideal composite pulses, with emphasis on application of three-level nitrogen-vacancy centers~\cite{Jing15}. It is shown that, for a three-level system, the effectiveness of LEOs does not depend on the details of the composite pulses but on the integral of the pulse sequence in the time domain. Recent studies show a significant advantage of a three-level system embedded in a triple quantum dot, which is associated with a decoherence-free subspace of a charge quadrupole qubit~\cite{Intro2}. The leakage errors are caused by noise and could be reduced by smoothly-varying short control pulses which are experimentally feasible. The system is modelled by two logical states and a leakage state coupled to one of them. %and has interesting applications in the triple quantum dot geometry which can generalize the concept of a DFS for semiconducting qubits~\cite{Intro}. 
Using perturbation technique and the quasistatic noise approximation, the leakage errors of single qubit operations can be suppressed by simple pulse sequences up to the sixth order in noise amplitude. While it is simple and efficient, the approach needs additional well-controlled pulses, %does not consider time-dependent noises or fluctuations~\cite{Flu2} 
and is only valid for small-amplitude noise. %resulting in a fundamental difficulty to apply error suppression methods under experimental conditions. 
These requirements may not be well satisfied during gate operations, especially when the strength and time-dependence of noise are not negligible in comparison with other control parameters. 
%In order to implement a high fidelity control of single qubit encoded in the triple quantum dot, an exact elimination of leakage errors is necessary. 

Here we present an exact solution to leakage elimination for a three-level system where the leakage state is coupled to one of the logical states, by using a simple sequence or circuit of experimentally-available finite rotations (gates).  The coupling strength between the logical state and the leakage state is assumed to be static during the operation time, which is experimentally feasible for semiconducting quantum dots~\cite{Intro2,Intro,PNAS}. This assumption may not be necessary as the later numerical simulation shows that our exact circuit performs perfectly, even in the presence of time-dependent noise. Moreover, we explain the parameter settings in our approach including estimation of noise strength based on the exact circuit. 

{\it The model.}---We start with a model Hamiltonian represented in the basis spanned by two logical states and one leakage state~\cite{Intro2},
\begin{equation}\label{Hami}
    H=H_{\rm{z}}+H_{\rm{x}}+H_{\rm{leak}},
\end{equation}
with

\begin{equation*}
    \begin{split}
    &H_{\rm{z}}=\frac{\epsilon_{\rm{q}}}{2}\left(
    \begin{array}{ccc}
        1 & 0 & 0 \\
        0 & -1 & 0 \\
        0 & 0 & -\zeta
    \end{array}
    \right),\
    H_{\rm{x}}=g\left(
    \begin{array}{ccc}
        0 & 1 & 0 \\
        1 & 0 & 0 \\
        0 & 0 & 0
    \end{array}
    \right),\ \rm{and} \\
    &H_{\rm{leak}}=\xi\left(
    \begin{array}{ccc}
        0 & 0 & 0 \\
        0 & 0 & 1 \\
        0 & 1 & 0 
    \end{array}
    \right),
    \end{split}
\end{equation*}
where we use the same notations as in Ref.~\cite{Intro2}. Here $\epsilon_{\rm{q}}$ and $g$ are independent control parameters for rotations with respect to the $z$ and $x$ directions. $H_{\rm{leak}}$ stands for a coupling between the leakage state and one of the logical states, and $\zeta$ is the scaled leakage state energy in the absence of coupling~\cite{Flu2,WS1,GH}. %For semiconducting charge quadrupole (CQ) qubits formed in three adjacent dots sharing one electron~\cite{Intro}, there happens to exist a Decoherence-Free space (DFS) against the uniform electric field fluctuations. 
A charge quadrupole (CQ) qubit is formed in three adjacent semiconducting quantum dots sharing a single electron and is embedded in the localized charge basis $\{|100\rangle,|010\rangle,|001\rangle\}$, where the basis states denote the electron being in the first, second or the third dot, respectively. The system Hamiltonian reads
\begin{equation}\label{HQC}
    H_{\rm{CQ}}=\left(
    \begin{array}{ccc}
        \epsilon_{\rm{d}} & t_{\rm{A}} & 0\\
        t_{\rm{A}} & \epsilon_{\rm{q}} & t_{\rm{B}}\\
        0 & t_{\rm{B}} & -\epsilon_{\rm{d}}
    \end{array}
    \right)+\frac{U_1+U_3}{2},
\end{equation}
\begin{figure}[htbp]
\centering
\includegraphics[width=3.1in]{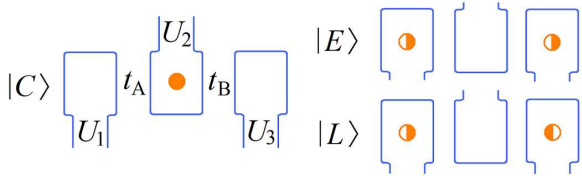}
\caption{Schematic diagram of CQ qubits. Adjacent quantum dots are represented in blue. $U_{1,2,3}$ are on-site potentials, and $t_{\rm{A}}$($t_{\rm{B}}$) is the coupling between the left(right) and middle dots. Logical states are denoted by $|C\rangle, |E\rangle$ and leakage state by $|L\rangle$. The orange dots are electrons, where the full filled means an electron and the half filled represents that electrons are in equal superposition states.}\label{F1}
\end{figure}where $U_{1,2,3}$ are the on-site potentials for the three dots. Here $t_{\rm{A,B}}$ are tunnel couplings between adjacent dots, and $\epsilon_{\rm{d}}=(U_1-U_3)/2$ ($\epsilon_{\rm{q}}=U_2-(U_1+U_3)/2$) denotes the dipolar (quadrupolar) detuning parameter. A new set of bases consisting of logical qubtis $|C\rangle,|E\rangle$ and a leakage state $|L\rangle$ is defined by~\cite{Intro2,Intro}
\begin{equation}\label{CEL}
    |C\rangle=|010\rangle,\ |E\rangle=\frac{|100\rangle+|001\rangle}{\sqrt{2}},\ |L\rangle=\frac{|100\rangle-|001\rangle}{\sqrt{2}},
\end{equation}
and a schematic diagram is presented in Fig.~(\ref{F1}). The Hamiltonian in the new basis is transformed into
\begin{equation}\label{THQC}
    \tilde{H}_{\rm{CQ}}=\left(
    \begin{array}{ccc}
    \frac{\epsilon_{\rm{q}}}{2} & \frac{t_{\rm{A}}+t_{\rm{B}}}{\sqrt{2}} & \frac{t_{\rm{A}}-t_{\rm{B}}}{\sqrt{2}} \\
    \frac{t_{\rm{A}}+ t_{\rm{B}}}{\sqrt{2}} & -\frac{\epsilon_{\rm{q}}}{2} & \epsilon_{\rm{d}} \\
    \frac{t_{\rm{A}}-t_{\rm{B}}}{\sqrt{2}} & \epsilon_{\rm{d}} & -\frac{\epsilon_{\rm{q}}}{2} 
    \end{array}
    \right),
\end{equation}
where a term proportional to the identity has been dropped. $\tilde{H}_{\rm{CQ}}$ is reduced to Eq.~(\ref{Hami}) under the conditions of $\zeta=1$, $\xi=\epsilon_{\rm{d}}$, and $g=(t_{\rm{A}}+t_{\rm{B}})/\sqrt{2}$. In case that $t_{\rm{A}}=t_{\rm{B}}$ and $\epsilon_{\rm{d}}=0$ are satisfied, $\tilde{H}_{\rm{CQ}}$ supports a decoherence-free subspace against uniform electric field fluctuations~\cite{Intro}.

In the triple quantum dot system, $\epsilon_{\rm{d}}$ corresponds to an average dipolar detuning control parameter. Although $\epsilon_{\rm{d}}$ is set to be zero, its fluctuation $\delta\epsilon_{\rm{d}}$ breaks the DFS and causes leakage. It has been shown that the fluctuation of quadrupolar detuning control parameter is smaller than $\delta\epsilon_{\rm{d}}$ and is thus neglected. Now we focus on the influence of $\delta\epsilon_{\rm{d}}$ on the CQ qubit operations. Noise spectrum of $\delta\epsilon_{\rm{d}}$ is dominated by low-frequency fluctuations which are slow in comparison with gate operations~\cite{PNAS}. Therefore $\delta\epsilon_{\rm{d}}$ is assumed to remain constant during a given gate operation~\cite{Intro2,Flu2}. As a result, unitary operators for $x$ and $z$ rotations can be given by
\begin{equation}\label{OU}
    \begin{split}
        & U_{\rm{x}}(g,\delta\epsilon_{\rm{d}},\theta)=\exp\{-i[H_{\rm{x}}(g)+H_{\rm{leak}}(\delta\epsilon_{\rm{q}})]\theta/2g\} \\
        & U_{\rm{z}}(\epsilon_{\rm{q}},\delta\epsilon_{\rm{d}},\varphi)=\exp\{-i[H_{\rm{z}}(\epsilon_{\rm{q}})+H_{\rm{leak}}(\delta\epsilon_{\rm{q}})]\varphi/\epsilon_{\rm{q}}\},
    \end{split}
\end{equation}
with arbitrary angles $\theta$ and $\phi$. In the bang-bang limit where the control pulses switch instantaneously between two values, the angles are associated with the corresponding bang-bang gate time intervals $t_z$ and $t_x$, which are $\theta=t_z(\epsilon_{\rm{q}}/\hbar)$, $\phi=t_x(2g/\hbar)$. As shown by Eq.~(\ref{OU}), rotation operators are obviously {\em polluted} by $\delta\epsilon_{\rm{d}}$. Below, we will explain our exact solution to this problem.

{\it A set of ${\it su}(2)$ generators, finite rotations, exact elimination of leakage }.--- To suppress the fluctuation $\delta\epsilon_{\rm{d}}$ in $U_{\rm{x}}(g,\delta\epsilon_{\rm{d}},\theta)$, we start with the following three matrices
\begin{equation*}
    \begin{split}
    &M_1=\left(
    \begin{array}{ccc}
        0 & 1 & 0 \\
        1 & 0 & 0 \\
        0 & 0 & 0
    \end{array}
    \right),\
    M_2=\left(
    \begin{array}{ccc}
        0 & 0 & 0 \\
        0 & 0 & 1 \\
        0 & 1 & 0
    \end{array}
    \right),\ \rm{and} \\
    &M_3=\left(
    \begin{array}{ccc}
        0 & 0 & -i \\
        0 & 0 & 0 \\
        i & 0 & 0 
    \end{array}
    \right).\   
    \end{split}
\end{equation*}
It can be shown that their commutation relations satisfy,
\begin{equation*}
    \begin{split}
        [M_1,M_2]=iM_3,~[M_2,M_3]=iM_1,~[M_3,M_1]=iM_2,
    \end{split}
\end{equation*}
indicating that these operators generate an ${\it su}(2)$ algebra. % It is well known that ${\it su}(2)$ plays an important role in the parametrization of rotations in three dimensional Euclidean space~\cite{AM}. 
An arbitrarily given finite rotation can be represented in an exponential form~\cite{AM}
\begin{figure*}[htbp]
\centering
\setcounter{equation}{8}
\begin{equation}\label{EQE}
    \begin{split}
    &\left(
    \begin{array}{ccc}
        \cos{\phi_2} & i\sin{\phi_2}\cos{\phi_3} & -\sin{\phi_2}\sin{\phi_3} \\
        i\cos{\phi_1}\sin{\phi_2} & \cos{\phi_1}\cos{\phi_2}\cos{\phi_3}-\sin{\phi_1}\sin{\phi_3} & i\cos{\phi_1}\cos{\phi_2}\sin{\phi_3}+i\sin{\phi_1}\cos{\phi_3} \\
        -\sin{\phi_1}\sin{\phi_2} & i\sin{\phi_1}\cos{\phi_2}\cos{\phi_3}+i\cos{\phi_1}\sin{\phi_3} & -\sin{\phi_1}\cos{\phi_2}\sin{\phi_3}+\cos{\phi_1}\cos{\phi_3}
    \end{array}
    \right)\\
    &~~~~~~~~~~~~~~~~~~~~~~~~~~~~=\left(
    \begin{array}{ccc}
        1+a^2(\cos{\alpha}-1) & ia\sin{\alpha} & ab(\cos{\alpha}-1) \\
        ia\sin{\alpha} & \cos{\alpha} & ib\sin{\alpha} \\
        ab(\cos{\alpha}-1) & ib\sin{\alpha} & 1+b^2(\cos{\alpha}-1)
    \end{array}
    \right)
    \end{split}
\end{equation}
\end{figure*}
\setcounter{equation}{5}
\begin{equation}\
    \exp[i(\gamma_1M_1+\gamma_2 M_2+\gamma_3 M_3)],
\end{equation}
where $\gamma_1$, $\gamma_2$, $\gamma_3$ are three continuous parameters and a linear combination of $M_i(i=1,2,3)$ indicates a specific rotation axis and the corresponding angle. On the other hand, the finite rotation can also be expressed by three Euler's angles $\phi_1$, $\phi_2$ and $\phi_3$,
\begin{equation}\label{Euler}
    \exp[i\phi_1 M_2]\exp[i\phi_2 M_1]\exp[i\phi_3 M_2],
\end{equation}
The relation between the two sets of parametrizations can be found by setting
\begin{equation}\label{MF}
    \begin{split}
        &\exp[i\phi_1 M_2]\exp[i\phi_2 M_1]\exp[i\phi_3 M_2]\\
        &=\exp[i(\gamma_1M_1+\gamma_2M_2+\gamma_3M_3)],
    \end{split}
\end{equation}
where the two sets $\phi_i$ and $\gamma_i$ ($i=1,2,3$) are in one-to-one correspondence. In our cases~(\ref{OU}), we can set $\gamma_1=\alpha a$, $\gamma_2=\alpha b$, $\gamma_3=0$, where $a^2+b^2=1$. It is easy to verify that the powers of $M_i$ satisfy
\begin{equation*}
    \begin{split}
        M_i^{2n}=I-\Delta(i),~ &M_i^{2n+1}=M_i,\\
        (aM_1+bM_2)^{2n}=&(aM_1+bM_2)^2,\\
        (aM_1+bM_2)^{2n+1}&=aM_1+bM_2,
    \end{split}    
\end{equation*}
where $n$ is a positive integer, $I$ is the three dimensional identity matrix, and $\Delta(i)$ is a matrix with $\Delta(i)_{ii}=1$, $\Delta(i)_{jk}=0$ for $j\ne i$ or $k\ne i$ $(j,k=1,2,3)$. Based on the above properties, we can derive an exact matrix equation~(\ref{EQE}) representing a system of nine nonlinear equations of which only three equations are independent. These independent equations determine $\phi_1=\phi_3$ and
\setcounter{equation}{9}
\begin{equation}\label{CD}
     \begin{split}
        \cos{\phi_2}=1+a^2&(\cos{\alpha}-1),\\
        \sin{\phi_2}\sin{\phi_1}=a&b(1-\cos{\alpha}),\\
        \sin{\phi_2}\cos{\phi_1}=&~a\sin{\alpha}.
     \end{split}
\end{equation}
Therefore angles $\phi_1$ and $\phi_2$ can be expressed in terms of $\alpha$, $a$ and $b$. Substituting Eq.~(\ref{MF}) to $U_{\rm{x}}(g,\delta\epsilon_{\rm{d}},\theta)$, we obtain
\begin{equation}\label{RXA}
    \begin{split}
        &U_{\rm{x}}(g,\delta\epsilon_{\rm{d}},\theta) \\
        &=\exp(i\beta_1H_{\rm{leak}})\exp(i\beta_2H_{\rm{x}})\exp(i\beta_1H_{\rm{leak}}),
    \end{split}
\end{equation}
with parameter constraints
\begin{equation}\label{CDS}
    \begin{split}
         a=-\theta/2\alpha,~&b=-\theta\delta\epsilon_{\rm{q}}/2g\alpha,\\
        \alpha=(\theta/2g)&\sqrt{g^2+\delta\epsilon_{\rm{d}}^2},\\
        \beta_1={\pm}\arcsin[ab(1&-\cos{\alpha})/\sin{\beta_2}]/\delta\epsilon_{\rm{d}},\\
        \beta_2={\pm}\arccos[1-&a^2(1-\cos{\alpha})]/g,
    \end{split}
\end{equation}
which are obtained from the first two Eqs.~(\ref{CD}). The sign of $\beta_1$ and $\beta_2$ can be further determined by checking the parameter solutions with the third Eq.~(\ref{CD}). By reversing Eq.~(\ref{RXA}) we obtain the ideal gate operator with respect to $x$ axis,
\begin{equation}\label{RX}
    \begin{split}
        &U_{\rm{ix}}(g,-2g\beta_2)=\exp[-iH_{\rm{x}}(g)(-2g\beta_2)/2g]\\
        &=\exp(-i\beta_1H_{\rm{leak}})U_{\rm{x}}(g,\delta\epsilon_{\rm{d}},\theta)\exp(-i\beta_1H_{\rm{leak}}).
    \end{split}
\end{equation}
and eliminate the leakage $H_{\rm{leak}}$. The unitary operator $\exp(-i\beta_1H_{\rm{leak}})$ is the special case of the imperfect gate $U_{\rm{x}}(0,\delta\epsilon_{\rm{d}},2g\beta_1)$ which is experimentally feasible by microwave pulses in semiconducting dots. %Now we come to discuss control of parameters in Eq.~(\ref{RX}).% especially $\delta\epsilon_{\rm{d}}$.

{\it Parameter settings}.--- For a CQ qubit under noise $\delta\epsilon_{\rm{d}}$, an ideal $x$ rotation with angle $-2g\beta_2$ is generated by experimental parameters $\beta_1$, $\theta$, $g$ and $\delta\epsilon_{\rm{d}}$ in terms of the constraints~(\ref{CDS}). In semiconducting quantum dots, gate operations are implemented by microwave pulses so that $\theta$ can be modulated by the pulse width, and $g$ is determined by tunnel couplings $t_{\rm{A,B}}$. The spectrum of the noise $\delta\epsilon_{\rm{d}}$ in range of 5 kHz to 1 MHz has been shown by Hahn echo curves~\cite{PNAS}. Here our derivation suggests a new perspective to look into the noise $\delta\epsilon_{\rm{d}}$. An estimation of $\delta\epsilon_{\rm{d}}$ can be done by following steps: (i) prepare an initial state, for example $|C\rangle$. (ii) perform the three operations on the right side of Eq.~(\ref{RX}) with given $g$, $\beta_1$ and $\theta$ which has no limitation. The resultant operation in the logical subspace is an ideal $x$ rotation. (iii) measure the output state, and then  $\beta_2$ can be given. (iv) substitute $g$, $\beta_1$, $\theta$ and the measurement result of $\beta_2$ to Eq.~(\ref{CDS}), then $\delta\epsilon_{\rm{d}}$ are estimated. In experiments on semiconducting quantum dots, state initialization and readout take about 4 ms to 5 ms, and state manipulation needs about 1 ms~\cite{PNAS}. Therefore our estimation is allowed to be performed and repeated for several times and an effective strength curve of $\delta\epsilon_{\rm{d}}$ in the time domain can be concluded. Based on the noise spectrum as shown in Ref.~\cite{PNAS}, the strength of $\delta\epsilon_{\rm{d}}$ oscillates and the effective strength curve shows a periodicity. As a result, an effective $\delta\epsilon_{\rm{d}}$ at a desired operation time and its periodic extension can be given by the curve. 

{\it An arbitrary rotation without leakage}.--- An arbitrary leakage-free gate can be generated by three ideal $x$ and $z$ rotations. While the ideal $x$ rotation is given by~(\ref{RX}), in what follows, we will first show how to generate the ideal $z$ rotation. Let us start with the experimentally available $U_{\rm{z}}(\epsilon_{\rm{q}},\delta\epsilon_{\rm{d}},\phi)$ in Eq.~(\ref{OU}).  By using the commutation relation $[H_{\rm{z}},H_{\rm{leak}}]=0$,  $U_{\rm{z}}$ can be simply  decomposed into %by Baker-Campbell-Hausdorff formula~\cite{NielsenBook},
\begin{equation}\label{BCH}
    \begin{split}
        &U_{\rm{z}}(\epsilon_{\rm{q}},\delta\epsilon_{\rm{d}},\varphi)\\
        &=\exp[-iH_{\rm{z}}(\epsilon_{\rm{q}})\varphi/\epsilon_{\rm{q}}]\exp[-iH_{\rm{leak}}(\delta\epsilon_{\rm{q}})\varphi/\epsilon_{\rm{q}}].
    \end{split}
\end{equation}
Consequently, the leakage-free $z$ rotation can be realized by 
\begin{equation}\label{RZ}
    \begin{split}
        &U_{\rm{iz}}(\epsilon_{\rm{q}},\varphi)=\exp[-iH_{\rm{z}}(\epsilon_{\rm{q}})\varphi/\epsilon_{\rm{q}}]\\
        &=U_{\rm{z}}(\epsilon_{\rm{q}},\delta\epsilon_{\rm{d}},\varphi)U_{\rm{z}}(0,\delta\epsilon_{\rm{d}},-{\varphi}'),
    \end{split}
\end{equation}
where ${\varphi}'=\varphi\epsilon_{\rm{q}}$. Eq.~(\ref{RZ}) shows that only two different gates with the same $\delta\epsilon_{\rm{d}}$ are needed for the implementation of an ideal $z$ rotation. It does not require the detail of $\delta\epsilon_{\rm{d}}$ as well.

\begin{figure}[htbp]
\centering
\includegraphics[width=3.1in]{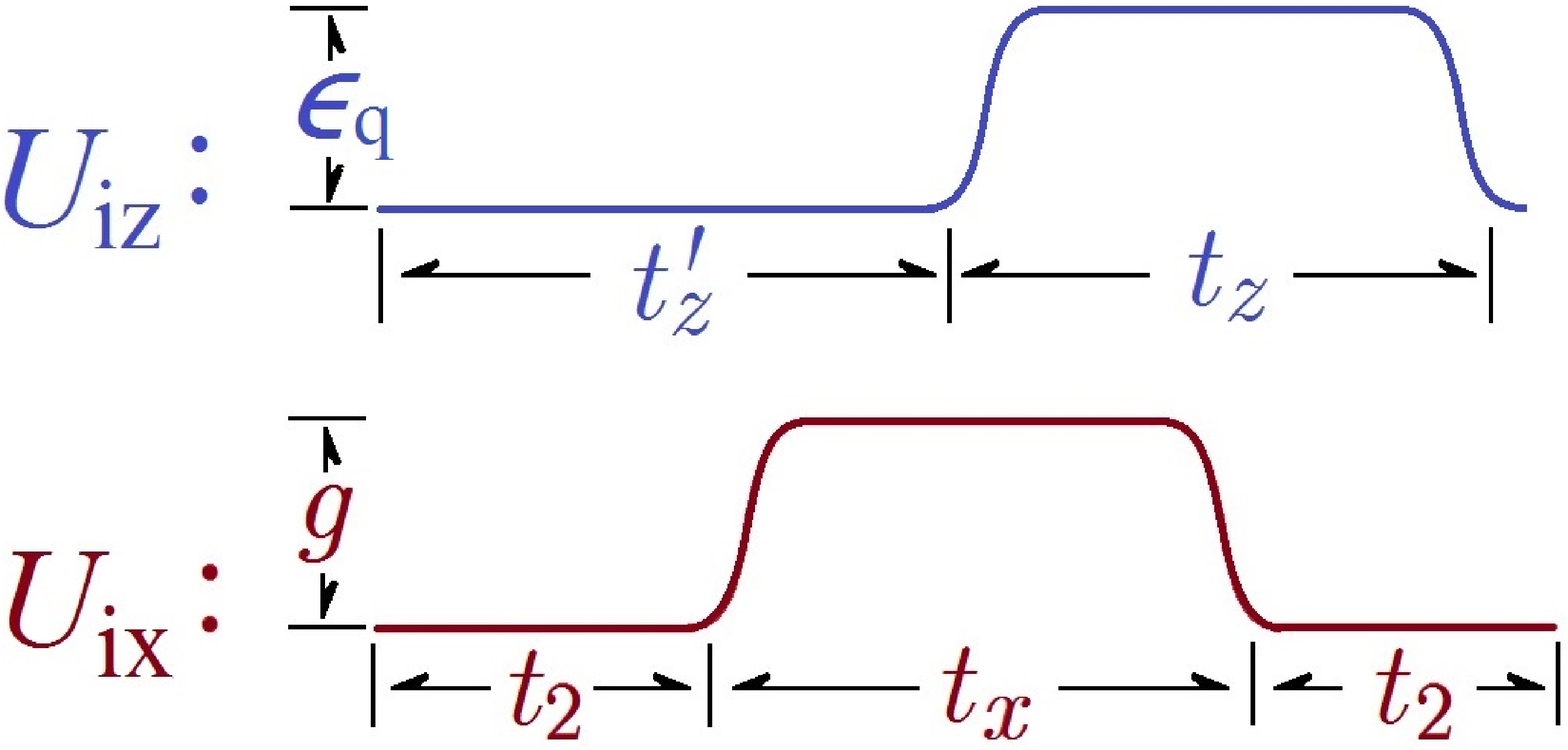}
\caption{Circuits for generating $U_{\rm{ix}}$ and $U_{\rm{iz}}$. The magnitude of applied pulses are $\epsilon_{\rm{q}}$ and $g$. The operation time $t_z$ ($t_z'$) are given through angle $\varphi$ ($\varphi'$) and detuning parameter $\epsilon_{\rm{q}}$, and $t_2$ and $t_x$ are given by angles $\beta_1$, $\beta_2$ and the coupling parameter $g$. The sign of angle parameters can be adjusted by setting the pulse width according to rotation period $2\pi$.} \label{F2}
\end{figure}

In general, it is well-known that an arbitrary leakage-free rotation for a single qubit can be implemented by combining $U_{\rm{ix}}$ and $U_{\rm{iz}}$, $i. e.$, three experimentally-available rotations for $x$ axis and two for $z$ axis, as sketched in Fig.~(\ref{F2}). %For $U_{\rm{ix}}$, a 0 pulse is applied for time $t_2$ and then a pulse of magnitude $g$ is switched on for $t_x$ followed by another 0 pulse for $t_2$. As to $U_{\rm{iz}}$ operation, a 0 pulse is applied for time $t_z'$ and then a pulse of magnitude $\epsilon_{\rm{d}}$ is switched on for $t_z$. 
%The magnitude and the width of pulses are determined by demanded rotating anglers, and all other parameters needed for pulses can be confirmed by above equations and methods. 

{\it Numerical Results}.--- %Now we show numerically the time evolution of a CQ qubit. 
Let us start with the initial state $|\psi_0\rangle=(|C\rangle+|E\rangle)/\sqrt{2}$, or its corresponding density matrix $\rho_0$
\begin{equation}\label{DOS}
     \begin{split}
    \rho_0=|\psi_0\rangle\langle\psi_0|=\left(
    \begin{array}{ccc}
        1/2 & 1/2 & 0 \\
        1/2 & 1/2 & 0 \\
        0 & 0 & 0
    \end{array}
    \right).
    \end{split}
\end{equation}
The time evolution of $\rho(t)$ by the propagator $U(t)$ is $\rho(t)=U(t)\rho_0 U(t)^{\dagger}$ for a noise channel. The dynamics of $\rho(t)$ will remain in the logical states and the matrix element  $\rho_{23}(t)=0$ if the propagator $U(t)$ does not contain leakage errors. If there are leakage errors in $U(t)$, where $\rho(t)$ is thus required to be an average over different noise channels denoted by $\delta\epsilon_{\rm{d}}$, we can use $\rho_{23}$ (or $\rho_{32}$) to characterize the errors.  %On the other hand, the noise term $\delta\epsilon_{\rm{d}}$ changes with time and is assumed to be constant during the operations which we focus on. Different $\delta\epsilon_{\rm{d}}$ could cause leakage errors with different magnitudes. As a result, the absolute value of the average of $\rho_{23}$ over a number of equal weighted $\delta\epsilon_{\rm{d}}$ , which is denoted by $|\rho_{23}\rangle_{\delta}|$, can represent the influence of the  noise changing on the quality of the operators.

\begin{figure}[htbp]
\centering
\includegraphics[width=3.3in]{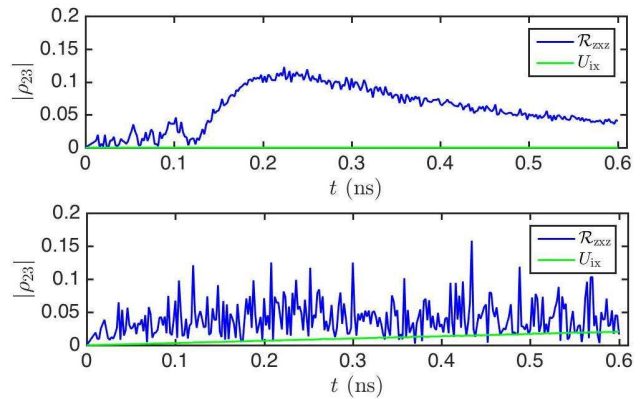}
\caption{$|\rho_{23}(t)|$ with $\mathcal{R}_{\rm{zxz}}$ (blue) and $U_{\rm{ix}}$ (green), respectively. The density matrix is averaged over random $\delta\epsilon_{\rm{d}}$.% in the top subfigure. %\red{and over 50 evolution processes in the bottom one}. \red{The parameters of $U_{\rm{ix}}$ in the bottom are solved by Eq.~(\ref{CD}) (or by Eq.~(\ref{CDS})) when ${\delta\epsilon}_{\rm{d}}$ is 0.3 GHz as the settings in Ref.~\cite{Intro2}}
%with the mean value $ \overline{\delta\epsilon}_{\rm{d}}=0.3~\rm{GHz}$ as in Ref.~\cite{Intro2}.
}
\label{F3}
\end{figure}

Figs.~{\ref{F3}} show the dynamics of $|\rho_{23}(t)|$ for the propagators $U_{\rm{ix}}$ and $\mathcal{R}_{\rm{zxz}}$, where the latter is defined as in Ref.~\cite{Intro2}. Note that we set parameters in $\mathcal{R}_{\rm{zxz}}$ such that it acts the same as $U_{\rm{ix}}$. The semiconducting dot experiments~\cite{Intro2, Flu} suggest that we can set $g/h$ to be 3.0 GHz
%$g/h\approx 3.0~{\rm{GHz}}$ 
and the evolution time from 0 ns to 0.6 ns. $\delta\epsilon_{\rm{d}}$ is a random number and is supposed to range from  0.2 GHz to 0.5 GHz. The top subfigure is the case when $\delta\epsilon_{\rm{d}}$ remains a constant for a given channel, where the gate $U_{\rm{ix}}$ is prefect since leakage is fully eliminated. The bottom illustrates the dynamics of $|\rho_{23}(t)|$ when $\delta\epsilon_{\rm{d}}$ is completely random over the course of time, $i. e.$,  the time-dependent noise. % changes with time and $U_{\rm{ix}}$ is implemented assuming $\delta\epsilon_{\rm{d}}=0.3~\rm{GHz}$.%We then calculate the time evolution of the amplitude $|\rho_{23}(t)|$ for the propagators $U_{\rm{x}}$ and $\mathcal{R}_{\rm{ZXZ}}$ defined in Ref.~\cite{Intro2}. % which are shown in Fig.~{\ref{F3}} together with the outcomes of $U_{\rm{ix}}$. From Fig.~{\ref{F3}}, we can see that the leakage error of our approach is always zero which is obviously lower than the former two especially when the evolution time increases.  
The two subfigures may imply two {\em bounds} of leakage effects, the time-independent bound and full time-dependent bound. The time-independent noise is more realistic as shwon in Ref.~\cite{PNAS}. It is noticeable that the present exact circuit performs perfectly even for full time-dependent noise.
%our exact circuit is insensitive to noise amplitudes.

{\it Conclusion}.--- We provide an exact solution to elimination of leakage errors in a three-level quantum model using simple circuits of gates. The model comprises of two logical states and a leakage state, which can be used to describe a triple quantum dot system supporting a DFS.  DFS is a well-known strategy in error suppression for quantum computation, which attracts significant attentions because of its minimal overhead requirements. %Our exact leakage-free circuits for a single qubit encoded in a DFS is an indispensable step towards are an indispensable step towards the implementation of fault-tolerant quantum computing with solid-state elements, and performs perfectly even in the presence of  time-dependent leakage. 
The concatenation of DFS and the exact circuits promises to give this approach a twofold resilience, against decoherence and stochastic leakage errors. Numerical simulation shows that  the exact circuits perform perfectly even in the presence of full time-dependent noise, indicating the stability and fault-tolerance of these circuits. Furthermore we propose an estimation of dipolar detuning control fluctuation to extract precise strength information of noise. The feasibility of our approach is ensured by the development of sophisticated experimental techniques~\cite{Intro2,Intro}. 

{\it Acknowledgments.}--- We acknowledge grant support from the Spanish MINECO/FEDER Grants FIS2015-69983-P, the Basque Government Grant IT986-16 and UPV/EHU UFI 11/55.

\end{document}